\documentclass[11pt,onecolumn,floatfix,amssymb,pra,aps,longbibliography,superscriptaddress,nopacs]{revtex4-2}
\usepackage{amssymb}
\usepackage{setspace}
\usepackage{amsmath}
\usepackage[dvips]{graphicx}
\usepackage{epsfig}
\usepackage{tabularx}
\usepackage{color}
\usepackage[pdfpagemode=UseNone,colorlinks=true,linkcolor=blue,citecolor=blue]{hyperref}

\begin{document}

\title[]{Probing gravitational interaction between milligram optomechanical oscillators operating in the quantum regime}

\author{M. Bonaldi}
\affiliation{Istituto Nazionale di Fisica Nucleare, TIFPA, 38123 Povo (TN), Italy}
\affiliation{{Institute of Materials for Electronics and Magnetism, Nanoscience-Trento-FBK Division, 38123 Povo (TN), Italy}}

\author{A. Borrielli}
\affiliation{Istituto Nazionale di Fisica Nucleare, TIFPA, 38123 Povo (TN), Italy}
\affiliation{{Institute of Materials for Electronics and Magnetism, Nanoscience-Trento-FBK Division, 38123 Povo (TN), Italy}}
\author{G. Di Giuseppe}
\affiliation{{Physics Division, School of Science and Technology, Universit\`a di Camerino, 62032 Camerino (MC), Italy}}
\affiliation{INFN, Sezione di Perugia, 06123, Perugia, Italy}
\author{N. Malossi}
\affiliation{{Physics Division, School of Science and Technology, Universit\`a di Camerino, 62032 Camerino (MC), Italy}}
\affiliation{INFN, Sezione di Perugia, 06123, Perugia, Italy}
\author{F. Marin}
\email[Corresponding author: ]{francesco.marin@unifi.it}
\affiliation{LENS, Via Carrara 1, 50019 Sesto Fiorentino (FI), Italy}
\affiliation{{Dipartimento di Fisica e Astronomia, Universit\`a di Firenze, Via Sansone 1, 50019 Sesto Fiorentino (FI), Italy}}
\affiliation{INFN, Sezione di Firenze, Via Sansone 1, 50019 Sesto Fiorentino (FI), Italy}
\affiliation{CNR-INO, L.go Enrico Fermi 6, 50125 Firenze, Italy}
\author{F. Marino}
\affiliation{INFN, Sezione di Firenze, Via Sansone 1, 50019 Sesto Fiorentino (FI), Italy}
\affiliation{CNR-INO, L.go Enrico Fermi 6, 50125 Firenze, Italy}
\author{F. Marzioni}
\affiliation{{Physics Division, School of Science and Technology, Universit\`a di Camerino, 62032 Camerino (MC), Italy}}
\affiliation{INFN, Sezione di Perugia, 06123, Perugia, Italy}
\author{R. Natali}
\affiliation{{Physics Division, School of Science and Technology, Universit\`a di Camerino, 62032 Camerino (MC), Italy}}
\affiliation{INFN, Sezione di Perugia, 06123, Perugia, Italy}
\author{G. Novelli}
\affiliation{{Physics Division, School of Science and Technology, Universit\`a di Camerino, 62032 Camerino (MC), Italy}}
\affiliation{INFN, Sezione di Perugia, 06123, Perugia, Italy}
\author{P. Piergentili}
\affiliation{{Physics Division, School of Science and Technology, Universit\`a di Camerino, 62032 Camerino (MC), Italy}}
\affiliation{INFN, Sezione di Perugia, 06123, Perugia, Italy}

\author{A. Pontin}
\affiliation{INFN, Sezione di Firenze, Via Sansone 1, 50019 Sesto Fiorentino (FI), Italy}
\affiliation{CNR-INO, L.go Enrico Fermi 6, 50125 Firenze, Italy}

\author{P. Sberna}
\affiliation{Tera-Hertz Sensing Group, Microelectronics Dept., Delft University of Technology, Delft, 2628 CD,  The Netherlands}

\author{E. Serra}
\affiliation{Istituto Nazionale di Fisica Nucleare, TIFPA, 38123 Povo (TN), Italy}
\affiliation{Tera-Hertz Sensing Group, Microelectronics Dept., Delft University of Technology, Delft, 2628 CD,  The Netherlands}
\author{G. Sordo}
\affiliation{SINTEF Digital, Dep. Smart sensors and microsystems, Gaustadalléen 23c, 0373 Oslo, Norway}
\author{D. Vitali}
\affiliation{{Physics Division, School of Science and Technology, Universit\`a di Camerino, 62032 Camerino (MC), Italy}}
\affiliation{INFN, Sezione di Perugia, 06123, Perugia, Italy}
\affiliation{CNR-INO, L.go Enrico Fermi 6, 50125 Firenze, Italy}

\begin{abstract}
We present an ongoing optomechanical experiment aimed at detecting gravitational interaction between milligram-scale oscillators, and exploring its interplay with quantum mechanics. The setup employs two microfabricated oscillators coupled via gravity and read out through a high-finesse optical cavity, at ultra-cryogenic temperature. Finite element simulations, noise modelling, and sensitivity estimates demonstrate that gravitational coupling can be detected with integration times ranging from minutes to hours. The experiment establishes a realistic platform for probing gravitational effects in systems approaching the quantum regime.
\end{abstract}

\maketitle

\section{Introduction}

Understanding whether gravity is fundamentally quantum remains one of the most profound open questions in physics. Although quantum mechanics successfully describes all known interactions at the microscopic scale, gravity has so far resisted a consistent quantum formulation. Direct tests at the Planck scale are experimentally inaccessible, motivating alternative approaches based on mesoscopic systems where quantum coherence and gravitational interaction may coexist \cite{Marshall2003,Bose2017,Marletto2017,Bose2025,Marletto2025}.

Mechanical systems operating in the quantum regime provide a promising platform to probe the interface between quantum mechanics and gravity. In particular, cavity optomechanics \cite{Aspelmeyer2014,Bowen2015} has enabled the preparation and control of non-classical motional states in micro- and nano-mechanical oscillators \cite{Teufel2011,Safavi-Naeini2012,Peterson2016,Arrangoiz-Arriola2019,Bild2023,Doeleman2023}.  However, a fundamental challenge remains: quantum control becomes increasingly difficult for large masses and low frequencies, which is the parameter regime where one could detect effects due to the gravitational interaction. Despite remarkable progress in quantum optomechanics, the mass scale of mechanical systems prepared in the quantum regime remains limited. The largest oscillators cooled to a phonon occupation below unity are bulk acoustic wave resonators with effective masses of approximately 0.5 mg \cite{Doeleman2023}. Reaching the quantum regime with substantially larger masses is therefore an important milestone, particularly for experiments aiming to investigate gravitational interactions in quantum mechanical systems.

At the opposite end of the parameter space, Westphal \textit{et al.}~\cite{Westphal2021} demonstrated gravitational coupling between $\sim0.1$ g  masses oscillators, at frequencies in the sub-Hz range. 
The thermal noise background is here much higher than the intrinsic quantum fluctuations. In order to reach this quantum level, some major improvements are envisaged, such as operating at higher frequency, yielding higher phonon energy and lower technical noise, at cryogenic temperature, and exploiting a cavity to achieve quantum limited readout sensitivity. 
Currently, a platform combining these setup characteristics with gravitationally coupled milligram-scale masses remains missing. An experiment aiming to address this gap is based on vibrating membranes embedded in microwave readout circuits \cite{Liu2021,Depellette2025}. A load realized with $\sim$mg gold spheres provides the gravitational interaction, and lowers the resonance frequency to $\sim 2\,$kHz. A different approach relies on high mechanical quality factor torsional oscillators, either with $\sim 1\,$mg mass and $\sim 20\,$Hz resonance frequency \cite{Agafonova2026}, or with $\sim 100\,$g mass and $\sim 10\,\mu$Hz resonance frequency \cite{Manley2026}.     

In this work, we present an ongoing experiment having the scope of detecting the gravitational interaction between a pair of twin mechanical oscillators in the quantum regime. More precisely, a coherent gravitational signal produced by an excited oscillator (Source) would be detected on a background close to the zero-point fluctuations of its twin oscillator (Probe). Signal and noise analysis shows that gravitational coupling between our milligram oscillators can be detected within realistic integration times, providing a concrete step toward testing gravity in regimes compatible with quantum optomechanics.

\section{Experimental concept} \label{sec: Experimental concept}

\begin{figure*}[hbt!]

\centering
\includegraphics[width=0.5\textwidth]{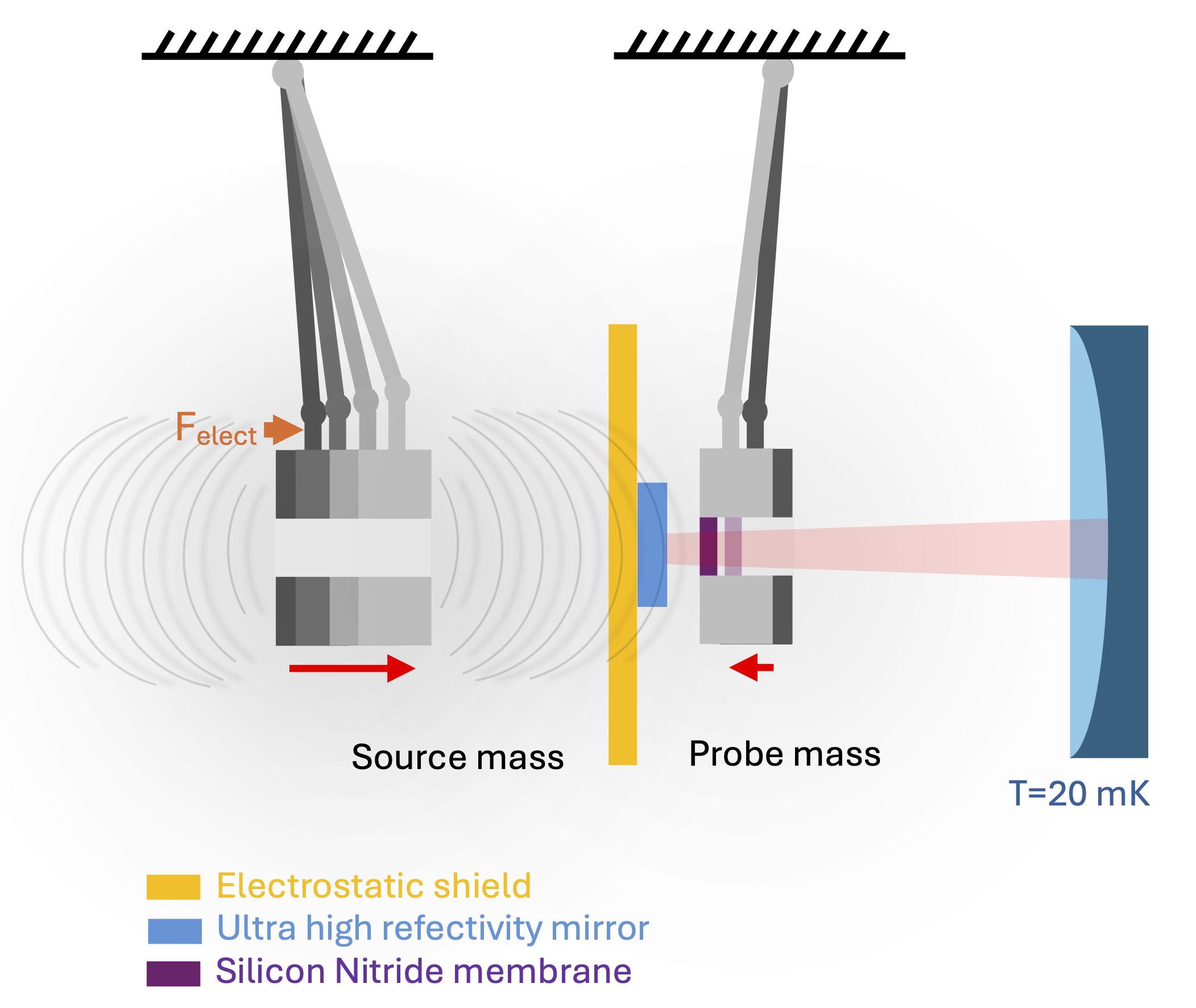}
\caption{Schematic of the proposed experiment. Two mechanical oscillators (Source and Probe) are positioned at sub-millimeter separations. The Source is driven at resonance, generating a time-dependent gravitational field. The Probe is coupled to a high-finesse optical cavity for displacement readout. \label{fig_setup}}
\end{figure*}

The system consists of a driven Source oscillator and a Probe oscillator acting as a force sensor. The two nominally identical devices are fabricated by micro-lithographic techniques on silicon wafers, with oscillating masses of approximately $\sim 20$ mg and resonance frequencies around $\sim 20$ kHz. The experiment operates in a regime where the gravitational coupling is modulated at the mechanical resonance frequency, enabling resonant amplification of the interaction signal. To achieve the required displacement sensitivity, the Probe is incorporated into an optical cavity (Fig. \ref{fig_setup}). The entire setup is maintained at ultra-cryogenic temperatures, in order to reduce thermal noise, a condition that also enhances the mechanical quality factor \cite{Marzioni2025}.

The experiment is structured in two main phases. In the first phase, the gravitational interaction between the Source and the Probe is measured: the Source oscillator is driven at resonance with an amplitude of approximately $\sim 1\,\mu$m, while the displacement of the Probe is read out via the optical cavity. In the second phase, the Probe oscillator is optically cooled down to a thermal occupancy of order unity through cavity readout combined with radiation-pressure-based active feedback cooling. Although this cooling does not improve the signal-to-noise ratio — since both signal and noise are reduced proportionally \cite{Pontin2014} — it allows the gravitational signal to emerge from a background dominated by the quantum fluctuations of the Probe, that produce specific and well identifiable signatures in the spectrum, such as asymmetric motional sidebands \cite{Safavi-Naeini2012}.

It is important to emphasize that, in this regime, the detectable gravitational signal still originates from the coherent motion of the Source rather than from its quantum fluctuations, and that the Source itself remains in a classical thermal state. A third phase, currently under feasibility investigation, aims at cooling the Source oscillator to its ground state as well, by means of a second optical cavity.

\section{Design of the mechanical oscillators}

\begin{figure*}[hbt!]
\centering
\includegraphics[width=1.0\textwidth]{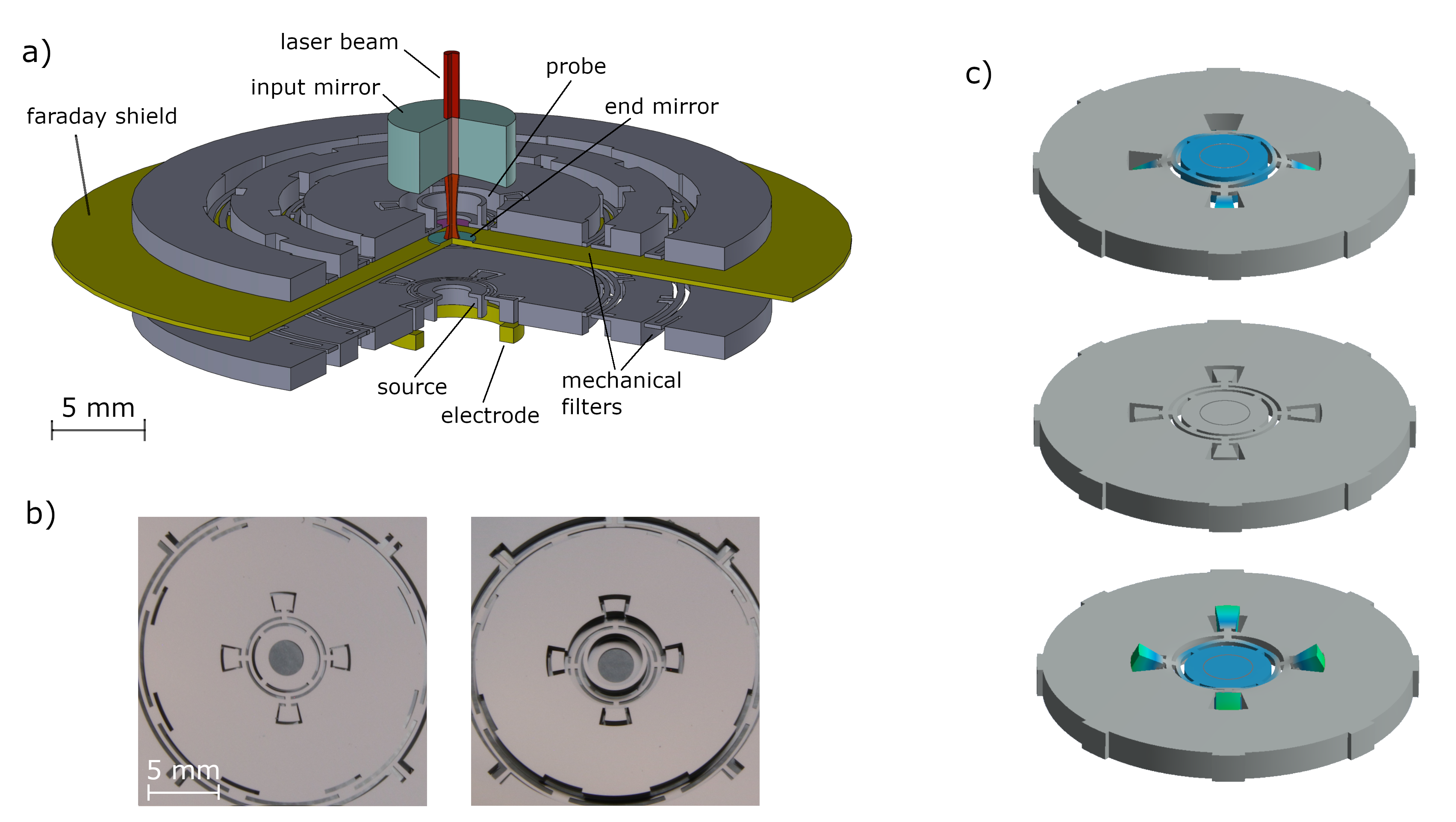}
\caption{a) Computed-aided design drawing of the mechanical oscillators (Source and Probe) assembled in the experimental configuration with a cavity readout and a Faraday shield. The Probe oscillator is equipped with the SiN membrane (shown in purple) to allow the coupling with the optical field. b) Front (left panel) and back (right panel) photos of a test device. c) Finite element simulation of the fundamental mechanical mode of the oscillator near $\sim 20$ kHz.\label{fig_oscillatori}}
\end{figure*}
  
The geometry of the mechanical oscillators, along with photographs of the fabricated devices, is presented in Fig. \ref{fig_oscillatori}. The primary moving element consists of a reinforced cylindrical shell with a diameter of about 4 mm, a thickness of 1.5 mm, and a mass of $23\,$mg. This disk is mechanically coupled to an additional larger disk, serving as an isolation mass, through a structure composed of alternating torsional and flexural spring elements arranged in series.
At the connection points, four counterweights are incorporated to balance the motion of the central mass (see the simulated mechanical mode shape in Fig. \ref{fig_oscillatori}c), ensuring that these junctions coincide with nodal points of the vibration. The counterweights of the Source and Probe oscillators are mounted with a 45-degree phase shift to avoid compensation of the gravitational forcing. The isolation mass is suspended from a further isolation ring, which is in turn connected to an external frame ring via appropriately designed beams. Additional silicon rings (not shown in the figure) are bonded to the isolation masses to increase their effective inertia.
The combined design of the spring geometry, nodal suspension, and isolation stages serves a dual purpose: minimizing energy transfer to the supporting frame in order to preserve a high mechanical quality factor, and providing strong mechanical decoupling between the Source and Probe at the resonance frequency.

The oscillator design is based on previously developed devices employed in optomechanical experiments \cite{Borrielli2014,Borrielli2015,Pontin2018}, here modified to increase the effective mass from approximately $\sim 10^{-7}$ kg to $\sim 10^{-5}$ kg. As a consequence, the resonance frequency is reduced from about $\sim 100$ kHz to $\sim 20$ kHz.
A key difference with respect to earlier implementations is the absence of a multilayer dielectric mirror. In prior designs, residual mechanical dissipation in the dielectric coating —despite the low strain achieved through optimized spring geometries \cite{Serra2012}— limited the mechanical quality factor to approximately $\sim 10^6$ at cryogenic temperatures ($\sim 5$ K).
In the present approach, coupling between the mechanical motion and the optical field in a cavity is achieved by the interaction with a suspended semi-transparent, 100 nm thick, SiN membrane. This configuration adds a very thin layer of material, in comparison with the 5 $\mu$m thick Bragg mirror, and enables the optical detection of the displacement by operating in the “membrane-in-the-middle” scheme \cite{Jayich2008}. It is worth noting, however, that membrane vibrational modes are not exploited in our experiment, as the pre-stressed membrane exhibits low compliance at the frequency of interest.

Previously fabricated circular membranes have demonstrated low optical absorption (imaginary part of the refractive index of $2 \times 10^{-6}$) and low surface roughness ($< 300$ pm) \cite{Serra2016}, making them well suited for use in high-finesse optical cavities with minimal thermal dissipation.
The target mechanical quality factor for the device is approximately 50$\times 10^6$, as estimated by finite element simulations using well-established modeling tools \cite{Serra2014}. The contribution to the total mechanical dissipation of each component of the oscillator is determined by its associated loss angle and its participation ratio in the motion of the normal mode under consideration. In this estimate, we take into account that structural dissipation in crystalline silicon allows for the realization of oscillators with characteristic dimensions of 10 mm and quality factors of 10$^8$ \cite{McGuigan78,Kleiman85}. At cryogenic temperatures, thermoelastic dissipation is negligible \cite{Zendri2008}.

The wafers used for device fabrication are monolithic crystalline silicon FZ (Floating Zone) 6-inch disks with a $\langle 100 \rangle$ orientation $\pm 1^\circ$, n-type doping, and high resistivity in the range of $5000$--$10000~\Omega\cdot\text{cm}$. They exhibit low bow and warp ($<40~\mu\text{m}$) and have a thickness of $1500 \pm 25~\mu\text{m}$. Since the fabrication process starts from a monolithic silicon wafer, unlike in Silicon-on-Insulator (SOI) technology, the buried oxide layer cannot be exploited as an etch-stop layer during Bosch-based bulk silicon micromachining. Consequently, the fabrication process requires a carefully optimized design strategy aimed at minimizing the number of etching steps, thereby ensuring a more uniform etched-area distribution and mitigating ARDE (Aspect-Ratio Dependent Etching) effects. To further improve process uniformity and dimensional control, double-sided etching with temporized process steps is employed. Two Probe and Source devices are fabricated on each 6-inch wafer. The fabrication process begin with the growth of a $300~\text{nm}$ thermal silicon oxide layer, which acts as an etch-stop layer for fabricating the silicon nitride (SiN) membrane in the $2.5~\text{mm}$-diameter opening of the Probe device. Subsequently, a $100~\text{nm}$ LPCVD SiN layer, characterized by a tensile stress of approximately $1~\text{GPa}$, is deposited. Following photoresist spin-coating, exposure, and development, the $250~\mu\text{m}$-thick spring structures is etched using deep reactive ion etching (DRIE) parameters specifically optimized to minimize sidewall tapering. After resist stripping, a  PE-CVD (Plasma Enhanced - Chemical Vapour Deposition)  step is performed with a 730 nm silicon oxide layer, which acts as an etch-stop layer for the subsequent backside DRIE process. A $4~\mu\text{m}$ PECVD oxide layer is then deposited on the backside of the wafer and used as a hard mask. After backside lithography, the exposed oxide regions is removed. An additional lithographic step employing a resist layer is carried out to define the central hole of the probe devices, followed by the removal of $250~\mu\text{m}$ of silicon to form the $2.5~\text{mm}$-diameter opening. After resist removal, the oxide mask is used to define the resonator test masses by etching the remaining $1250~\mu\text{m}$ of silicon using the Bosch DRIE process. Following Tepla plasma cleaning for resist removal, vapor HF etching is employed to release the membrane and remove the residual oxide layers. Finally, the individual devices are manually separated from the wafer by breaking the four anchor beams supporting each structure. Front and back photos of a test device are shown in Fig. \ref{fig_oscillatori}b.

To drive the Source motion, we plan to employ an electrode system \cite{Borrielli2011}, although a method based on the deposition of piezoelectric materials will be evaluated as a possible alternative. As shown in Fig. \ref{fig_oscillatori}a, the electrodes will be positioned above the four counterweights of the Source, at a distance of approximately 100 $\mu$m, and anchored to the device structure. A properly designed metallic layer keeps the counterweights at ground potential, while a bias voltage modulated at the oscillator frequency is applied to the electrodes. Given the presence of the conductive films necessary for voltage control, the design target for the Source quality factor is limited to 1$\times 10^6$. We point out that this configuration enables the application of an internally compensated force within the Source structure, relaxing the requirements for mechanical cross-talk rejection between the two oscillators, as it will be discussed in Sec. \ref{sec:Spurious}. The target oscillation amplitude for the Source is 1$\,\mu$m, achievable with a static voltage of 100 V and a 10\% modulation. In this case, the amplitude of the oscillating force on each counterweight, with a surface area of 1.05 mm$^2$, is approximately 0.25$\,\mu$N. 
We observe that the drive frequency must be tuned to the Probe's resonance to exploit the gravitational signal amplification factor, but also to the Source's resonance, which would otherwise be unable to reach the target oscillation amplitude. Therefore, it is necessary that the frequencies of the two oscillators match within $\pm 0.5\,{\rm Hz}$. This requirement exceeds the average reproducibility of the microfabrication techniques employed; therefore, after selecting a Probe/Source pair with suitable frequencies, it will be necessary to implement a tuning process involving sequential gold film depositions on the Source. To our knowledge, such a procedure has never been demonstrated and may prove challenging, given that the tuning must be maintained at cryogenic temperatures.

\section{Experimental setup}

\begin{figure*}[hbt!]
\centering
\includegraphics[width=.75\textwidth]{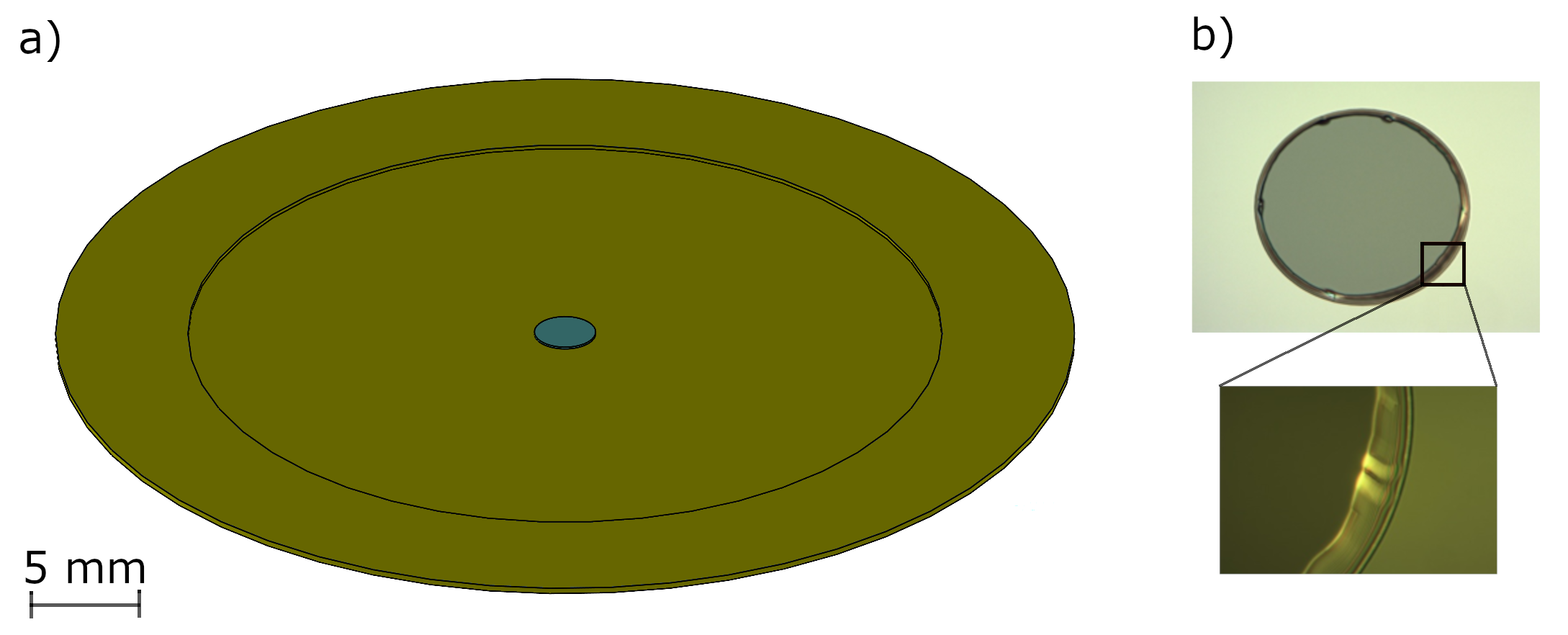}
\caption{a) 
Drawing of the Faraday shield with the high-reflectivity mirror. The central deposition, with thickness $6.67~\mu\text{m}$, and the spacer on the outer edge with thickness $20~\mu\text{m}$, are visible. These elements have been shown with exaggerated thickness in the figure for clarity. b) Photo and detail of the central mirror with $2.5~\text{mm}$ diameter.
\label{Fig_faraday}}
\end{figure*}

The distance between the Source and Probe oscillators is $550\, \mu$m. A bifunctional element (shown in Fig. \ref{Fig_faraday}), based on a silicon disk of thickness $300 \,\mu$m and diameter 50 mm, is placed between them. The surface facing the Source is coated with a gold layer 
%of thickness 200 nm, 
providing electrostatic shielding that suppresses electric Source–Probe coupling \cite{Westphal2021}. On the opposite surface, a dielectric multilayer disk of diameter 2.5 mm is deposited at the center, forming the end mirror of the optical cavity. The bifunctional element is mounted in direct contact with the frame of the Probe oscillator, such that their spacing and parallelism are defined by silicon micromachining.

Fabrication of the bifunctional element starts from a SOI wafer. The $22.5~\mu\text{m}$ device layer is completely etched to form a 33.6 mm-diameter circular region using Bosch DRIE, together with a 50 mm-diameter circular scribe line. The resulting annular edge region, with a thickness of $22.5~\mu\text{m}$, serves as the mirror spacer. The $500~\text{nm}$ buried oxide is removed before deposition of the high-reflective coating onto the remaining $300~\mu\text{m}$ handle layer. The multilayer coating is deposited by ion-beam sputtering and consists of 41 alternating $\text{Ta}_2\text{O}_5/\text{SiO}_2$ layers, with a total thickness of $6.67~\mu\text{m}$. It provides a nominal reflectivity of $99.999\%$ at $1064~\text{nm}$ (Nd:YAG wavelength).  To prevent substrate curvature and ensure optimal mirror planarity, a circular coating area with a diameter of $2.5~\text{mm}$ is defined by resist lift-off, taking particular care in compensating stresses in the SOI wafer through symmetric removal of the oxide on the SOI back side. As a final step, a low-stress, electron-beam-evaporated Au film with a thickness of $200~\text{nm}$ is deposited on the backside of the SOI wafer to provide the required electrostatic shielding.

\begin{figure*}[hbt!]
\centering
\includegraphics[width=1.0\textwidth]{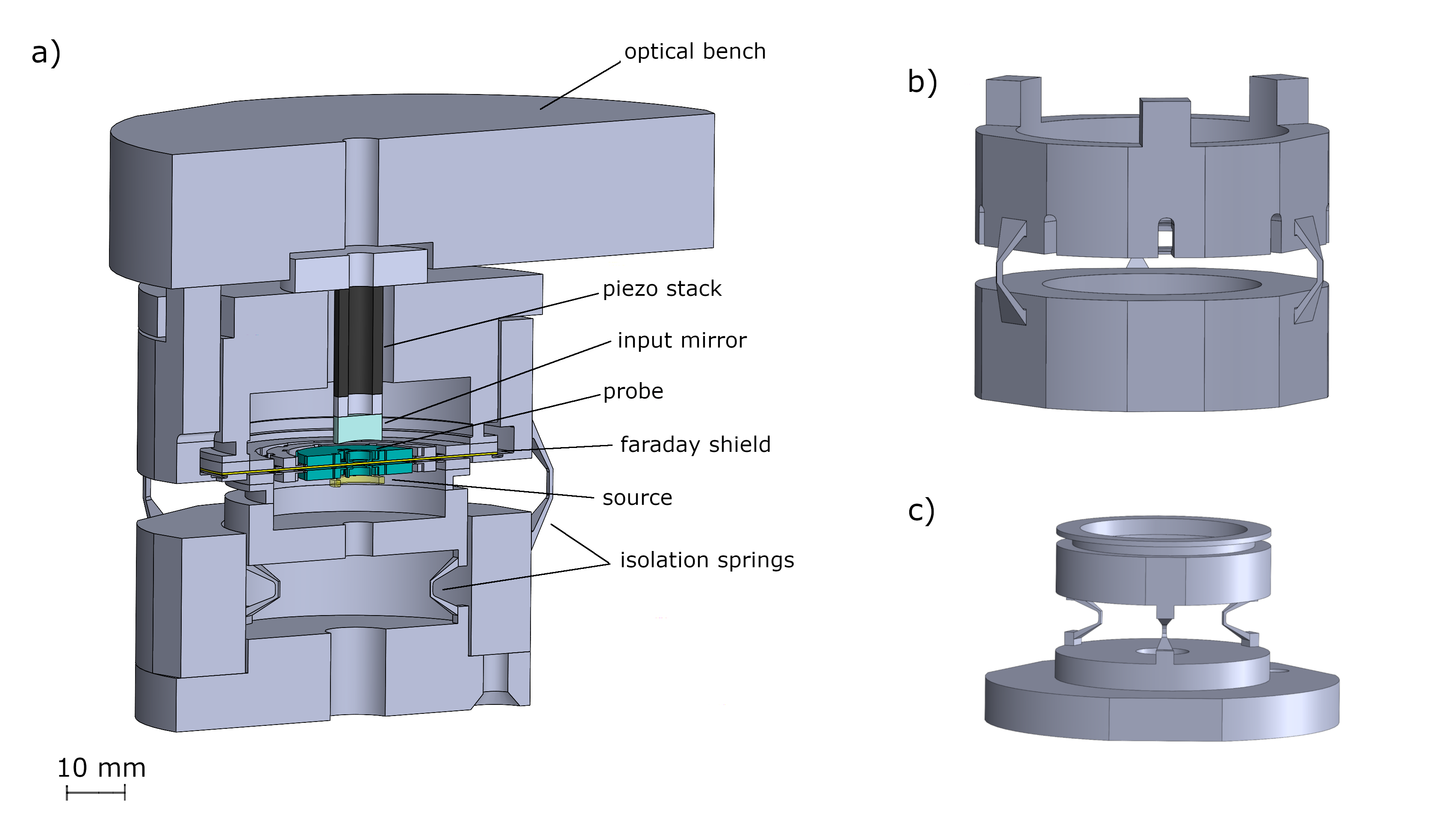}
\caption{
a) The Source and Probe devices are housed in an Invar structure supported by an optical bench, on which lenses and beam-steering stages (not shown in the picture) are fixed. Elastic elements between the different parts are used to reduce mechanical cross-talk between the Source and Probe oscillations. In total, although facing each other at a distance of approximately 0.5~mm, the Source and Probe are separated by two stages of mechanical suspension.
b) Detail of the housing structure.
c) Detail of the Source support.
\label{fig_cavita}}
\end{figure*}

The oscillators and the bifunctional element are housed in a monolithic Invar structure (Fig. \ref{fig_cavita}) optimized using Finite Element Method (FEM) simulations. The design incorporates an oscillator clamping system that provides sufficient elasticity to accommodate differential thermal contraction during cooling, thereby preventing fracture of the silicon components. The Source and Probe chambers are connected by a two-stage C-spring system which provides mechanical isolation of 60 dB at the resonance frequency of 20 kHz. The Probe chamber also hosts the input mirror of the optical cavity and is mounted vertically beneath the optical bench. The latter consists of a 20 mm thick Invar plate supporting the optical components for beam delivery, shaping, and alignment.

\begin{figure*}[hbt!]

\centering
\includegraphics[width=1.0\textwidth]{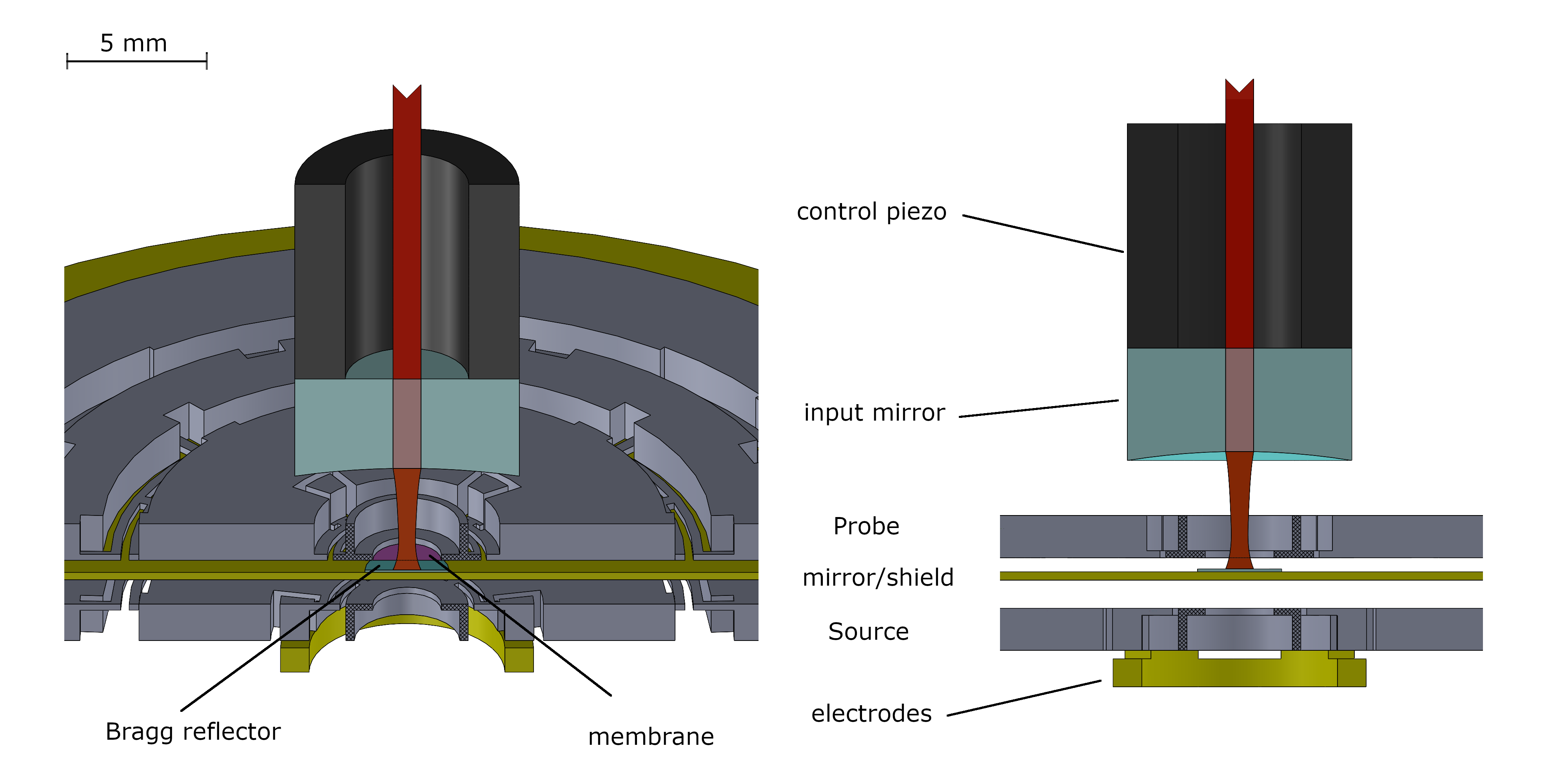}
\caption{
Cross-sectional details of the oscillators within the optical cavity setup. While the various components are drawn to scale, the distances between them have been exaggerated in the figure for clarity. The Source and Probe sections involved in the gravitational interaction study are hatched.
\label{fig_cavity_zoom}}
\end{figure*}

The Probe displacement is read out using a 3 mm long optical cavity, formed by a concave input coupler bonded to a piezoelectric transducer, and the high reflectivity mirror integrated on the bifunctional element (see Fig. \ref{fig_cavity_zoom}). As in our previous cryogenic optomechanical experiment employing the “membrane-in-the-middle” configuration \cite{Chowdhury2019}, the cavity is operated in a strongly overcoupled regime and exhibits a finesse of 20000.

The signal and local-oscillator beams are combined into a single-mode fiber and delivered to the cryogenic optical bench, which is mounted on the last stage of the mechanical suspension inside the cryostat. The bench hosts the alignment and mode-matching optics, as well as the cavity assembly. Light reflected from the cavity, carrying the signal and local oscillator, is collected into a multimode fiber and routed outside the cryostat for heterodyne detection. Using a multimode fiber for the output simplifies the alignment of the reflected beam and improves collection efficiency. The shared optical path of the signal and local oscillator outside the cavity provides efficient rejection of relative phase noise. We have verified that the use of a multimode fiber on the output path does not degrade this rejection, or introduce additional phase noise at the frequency of interest. The heterodyne signal is used both to lock the laser frequency to a cavity resonance and to infer the Probe displacement.

\begin{figure*}[hbt!]
\centering
\includegraphics[width=0.75\textwidth]{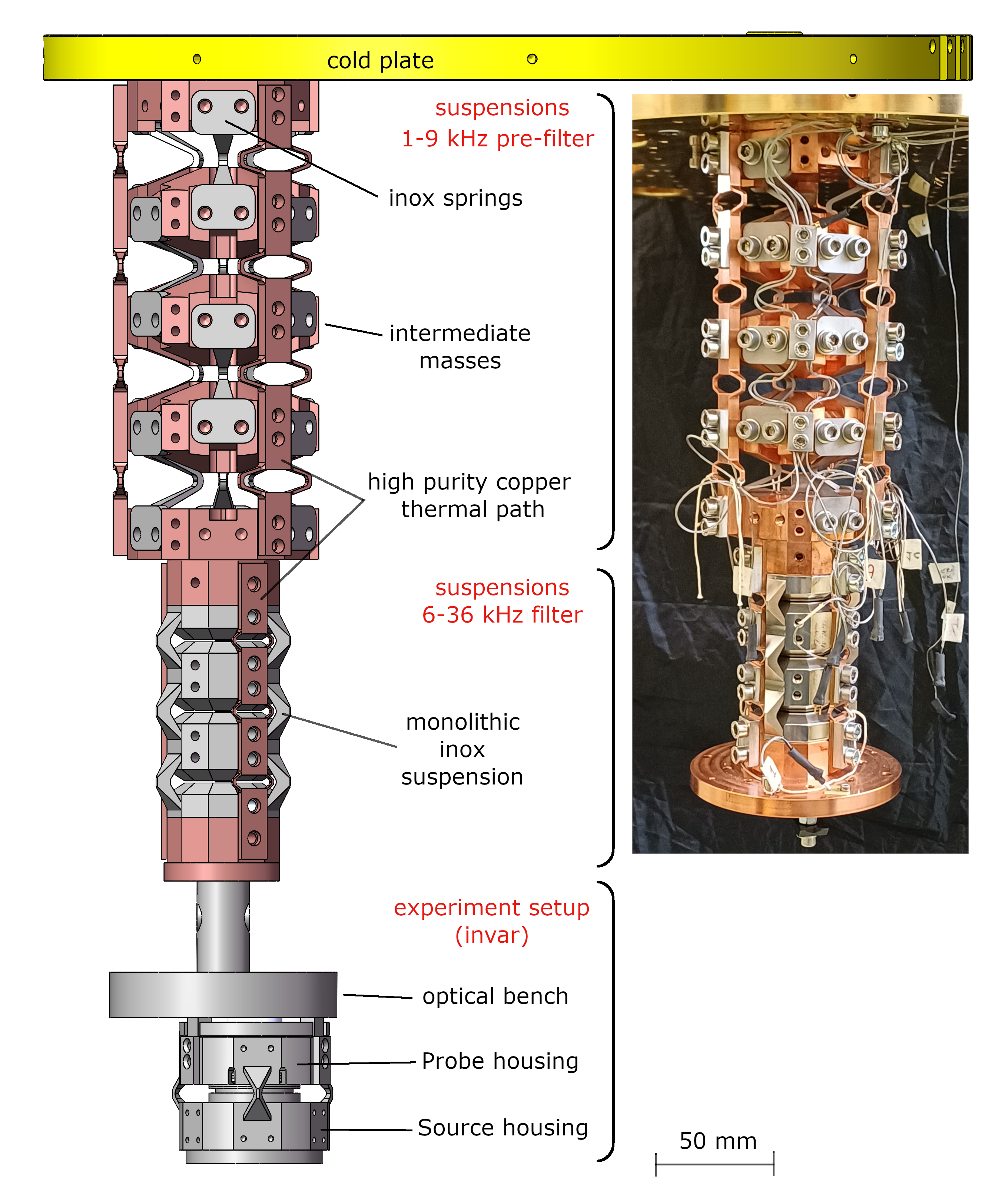}
\caption{CAD drawing of the complete measurement apparatus suspended from the cold plate of the dilution refrigerator. The two isolation stages are highlighted: the 1–9 kHz pre-filter and the 6–36 kHz filter, featuring copper thermal links separated from the steel suspension springs. The thermal links do not significantly modify the mechanics, i.e., they do not consistently stiffen the suspension chain, and are made of high-purity copper (99.999\%) with a residual resistance ratio (RRR) of approximately 2000. The optical bench and cavity structure are supported by the suspension chain. Thermal links for the cavity are planned, but not shown in the drawing. The photo on the right shows the suspension stages installed in the dilution refrigerator.\label{fig_sospensioni}}
\end{figure*}

The cryostat is a pulse tube precooled dilution refrigerator, customized from a Leiden Cryogenics CF-900-CS81, with a cooling power of about $100\,\mu$W at 45 mK. 
It contains two multi-stage suspension systems, where the first one is a pre-filter working in the 1-9 kHz range, and the second one is efficient in the 6-36 kHz range. Both suspensions, shown in Fig. \ref{fig_sospensioni}, consist of a cascade of spring-mass systems, including structural steel springs and soft copper springs as thermal links. The high frequency stage is fabricated from a single steel block. The overall reduction of the vibrations of the suspension point of the optomechanical cavity is better than $50\;$dB.

\section{Noise budget and sensitivity}

\subsection{Probe oscillator: thermal noise and susceptibility}

\begin{figure}[hbt!]
\centering
\includegraphics[width=85mm]{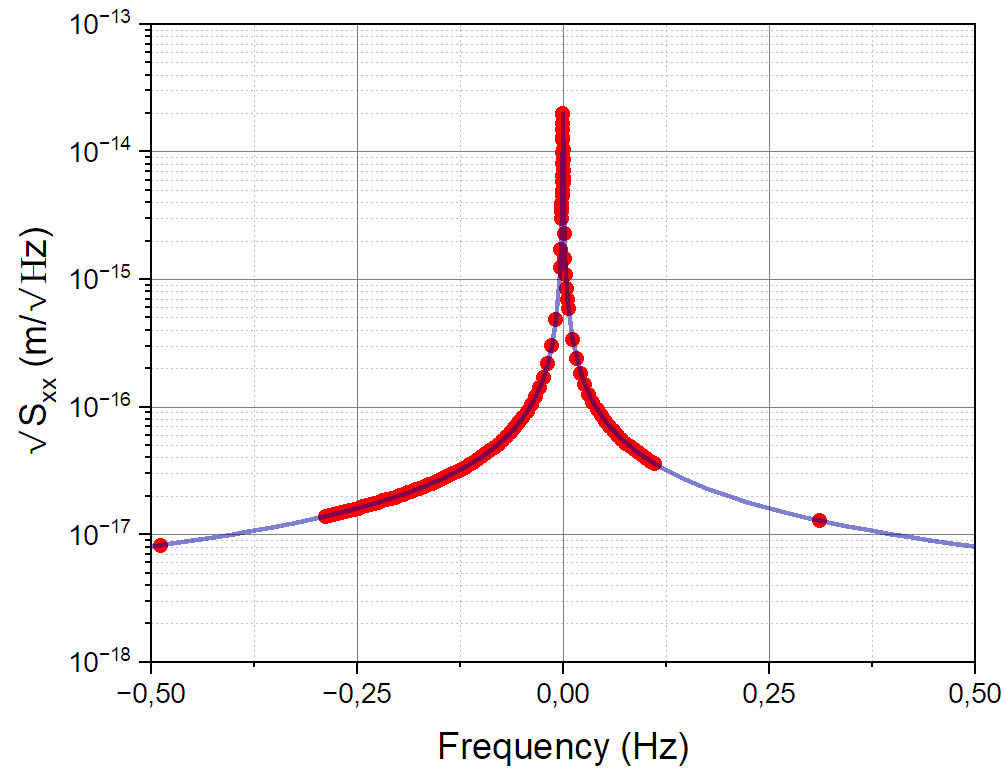}
\caption{Displacement noise spectrum of the Probe oscillator, centered around the resonant frequency $\Omega/2\pi\simeq19.5\,$kHz. Points correspond to finite-element simulation results, while the dashed line shows the fit used to evaluate the quality factor and the effective mass of the oscillator.
\label{fig_thermal}}
\end{figure}

Even at the target cryogenic temperature $T=20$ mK, thermal noise is expected to provide the dominant contribution to the noise of the Probe oscillator. Its spectral density, shown in Fig. \ref{fig_thermal}, has been evaluated \cite{Levin1998} using FEM, assuming a readout at the center of the SiN membrane on a circular area with a diameter of 100 $\mu$m. In the simulations, all components of the setup are considered, namely oscillators, mechanical filters, the SiN membrane, electrodes, etc., each contributing its own structural dissipation at cryogenic temperatures. Fitting the thermal peak with the response function of a single mechanical mode yields a resonance frequency $\Omega/2\pi \simeq19.5\,$kHz, a quality factor of $48 \times 10^6$, and an effective mass of 73 mg. This mass includes both the cylindrical shell, whose gravitational coupling is being studied, and the four counterweights. We stress once again that in the experimental setup, the Source and Probe devices are mounted rotated by 45 degrees, such that the gravitational interaction between the counterweights can be neglected. The physical mass of the central  reinforced cylindrical shell is $23\,$mg. The spectral density is $ 3\times 10^{-14}\, \mathrm{m/\sqrt{Hz}} $ at resonance, and decreases down to $ 10^{-17} \,\mathrm{m/\sqrt{Hz}} $ for a detuning of $ \pm\, 0.5 $ Hz. From the spectrum shown in Fig. \ref{fig_thermal}, we verify that the Probe oscillator can reach the target quality factor and extract the parameters that determine its response to the oscillating gravitational force.

\begin{figure}[hbt!]
\centering
\includegraphics[width=85mm]{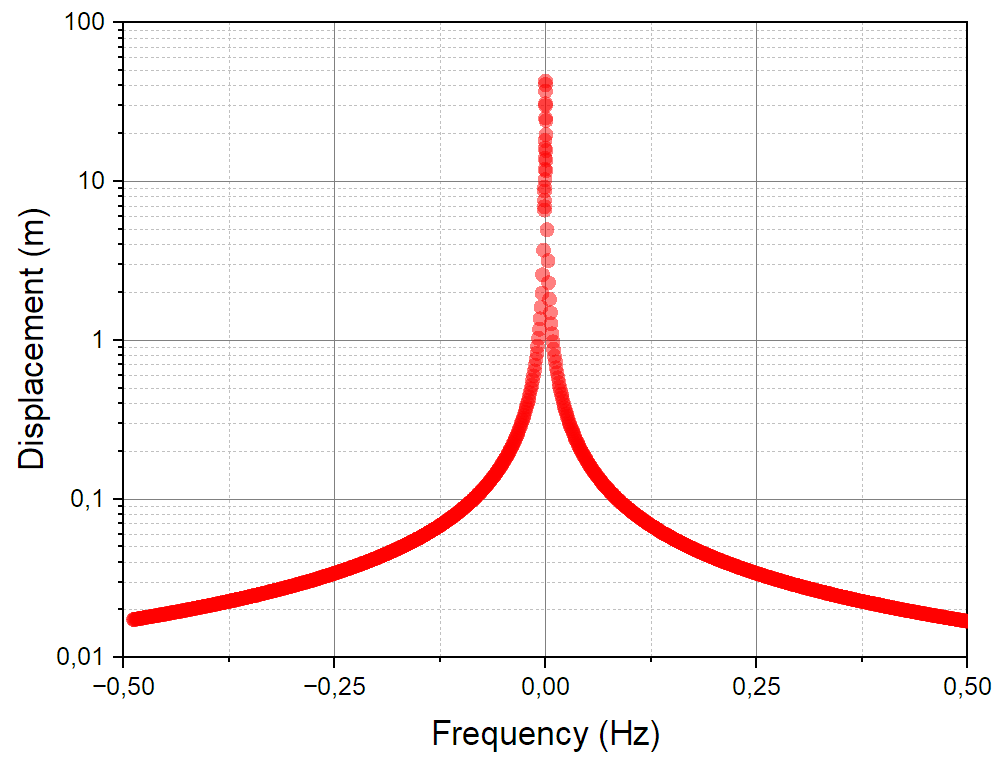}
\caption{Oscillation amplitude read out at the center of the membrane when the system is forced by an oscillating force of amplitude $1\,$N applied homogeneously over the volume of the central reinforced cylindrical shell. Points correspond to finite-element simulation results. The center frequency is $\Omega/2\pi\simeq19.5\,$kHz.
\label{fig_TF}}
\end{figure}

The same FEM model also provides the response to a force applied rigidly on the central cylindrical mass, useful for evaluating the displacement induced by the gravitational interaction signal and determining the necessary readout sensitivity. In Fig. \ref{fig_TF} is reported the oscillation amplitude at the center of the membrane for an oscillating force of amplitude $1\,$N applied homogeneously over the volume of the central reinforced cylindrical shell. 

Owing to the low compliance of the pre-stressed membrane at the frequency of interest,
the resulting displacement is only weakly dependent on the details of the distribution of an axially symmetric force. Therefore, this calculation provides a good approximation of the Probe response to the gravitational forcing generated by the Source. The amplitude at resonance is 42 m/N.

The dimensions of the oscillators and the parameters required for the preliminary sensitivity analysis are listed in Tab. \ref{tab:parametri}. From these values, the fluctuation-dissipation theorem allows us to derive the (single-side) power spectral density of the white thermal driving force acting on the oscillator as:
\begin{equation}
\label{thermal_force_PSD}
S_{\mathrm{th}}=4\, k_B\, T\; \frac{m_{\mathrm{eff}} \,\Omega}{Q} \simeq 2.0\times 10^{-31}\,\frac{\mathrm{N}^2}{\mathrm{Hz}}
\end{equation} 
This value will be compared with the applied gravitational force signal and will be considered in determining the most appropriate measurement strategy.

\begin{table*}[h!]
  \begin{center}
    \begin{tabular}{l|c|c}  
    \hline
    physical mass for GW interaction (mg)  &  $m_{\rm g}$ & 23 \\
effective mass (mg) & $m_{\rm eff}$ & 73 \\
Probe mechanical quality factor & $Q_{\rm p}$  & 48$\times 10^{6}$\\
Source mechanical quality factor & $Q_{\rm s}$ & 1$\times 10^6$\\
resonant frequency (kHz) &  $\Omega/2\pi$ & 19.5 \\
temperature (K) & $T$ & 0.02\\
Source-Probe target distance (mm) & $d$ & 0.55 \\
SiN membrane diameter (mm) & $s$ & 2.45 \\
SiN membrane internal stress(GPa) & $\sigma_{\rm m}$ & 0.8 \\
\hline     
    \end{tabular}
  \end{center}
      \caption{Oscillator measurements and parameters useful for preliminary sensitivity estimation. Although the physical measurements of the devices may change slightly during the tuning of the manufacturing processes, these parameters should remain valid within 10\%. The Probe and Source oscillators have the same physical and effective masses.}
            \label{tab:parametri}
\end{table*}

\subsection{Readout sensitivity}

The useful measurement bandwidth is defined 
as the frequency range over which the force-noise-driven resonance peak in the displacement spectrum rises above the flat background set by the measurement imprecision.
Although the high mechanical quality factor makes the thermal-noise peak readily resolvable, achieving a reasonably large measurement bandwidth requires the high sensitivity provided by a high-finesse optical cavity. 

We identify two levels of imprecision required for the readout. The first pertains to the initial phase of the experiment and is set by the requirement of achieving a measurement bandwidth of at least 1 Hz. The thermal noise spectrum discussed above indicates that the corresponding imprecision must be $10^{-17} \,\mathrm{m/\sqrt{Hz}}$. The second level, relevant for the second phase of the experiment, is the imprecision required to cool the Probe oscillator to a phonon occupancy close to unity via feedback cooling. Using the parameters reported in Table I, the oscillator initially has an occupation number of $\sim$20,000 and a linewidth of 0.4 mHz. Upon cooling, it reaches an effective linewidth of $\sim 10$ Hz, implying a required readout sensitivity of $10^{-18}\, \mathrm{m/\sqrt{Hz}}$. This noise floor has already been achieved in cryogenic optomechanics experiments with cavity finesse comparable to that envisaged here, at frequencies in the 100–200 kHz range \cite{Pontin2018}.

A shot-noise limited heterodyne detection achieves the required sensitivities for a signal laser power of $\sim 0.1\,\mu$W in the first phase of the experiment, and $\sim 10\,\mu$W in the second phase. Even if a significant fraction of this power is extracted from the cryostat, maintaining the required temperature of the oscillator entails a heat removal rate that can be nontrivial at ultracryogenic temperatures. It is worth noting that we will use a data-driven approach to develop the details of the thermalization system.

Displacement is converted into laser frequency variations via the factor $\,\mathrm{d} \nu/\mathrm{d} x = \mathcal{R} (\nu_{\mathrm{L}}/L_{\mathrm{cav}})\,$  where $\nu_{\mathrm{L}}$ is the laser frequency, $ L_{\mathrm{cav}}$ is the cavity length, and $\mathcal{R} $ is a coefficient that depends on the coupling between the membrane position and the cavity field \cite{Jayich2008}. This coupling can be optimized by tuning the laser wavelength, achieving a value of $\mathcal{R} \simeq 1$. For the cavity envisaged for this experiment, this yields a conversion factor of $\sim10^{17}\, \mathrm{Hz/m}$, and the upper limits to the laser frequency noise are therefore $1\, \mathrm{Hz/\sqrt{Hz}}$ and $0.1\, \mathrm{Hz/\sqrt{Hz}}$ for the first and second phase of the experiment, respectively. The first level is met by commercial Nd:YAG NPRO lasers, while the second can be achieved via active stabilization to a reference cavity \cite{Bondu1996,Conti2003}.

\subsection{Gravitational signal}
\begin{figure}[hbt!]
\centering
\includegraphics[width=85mm]{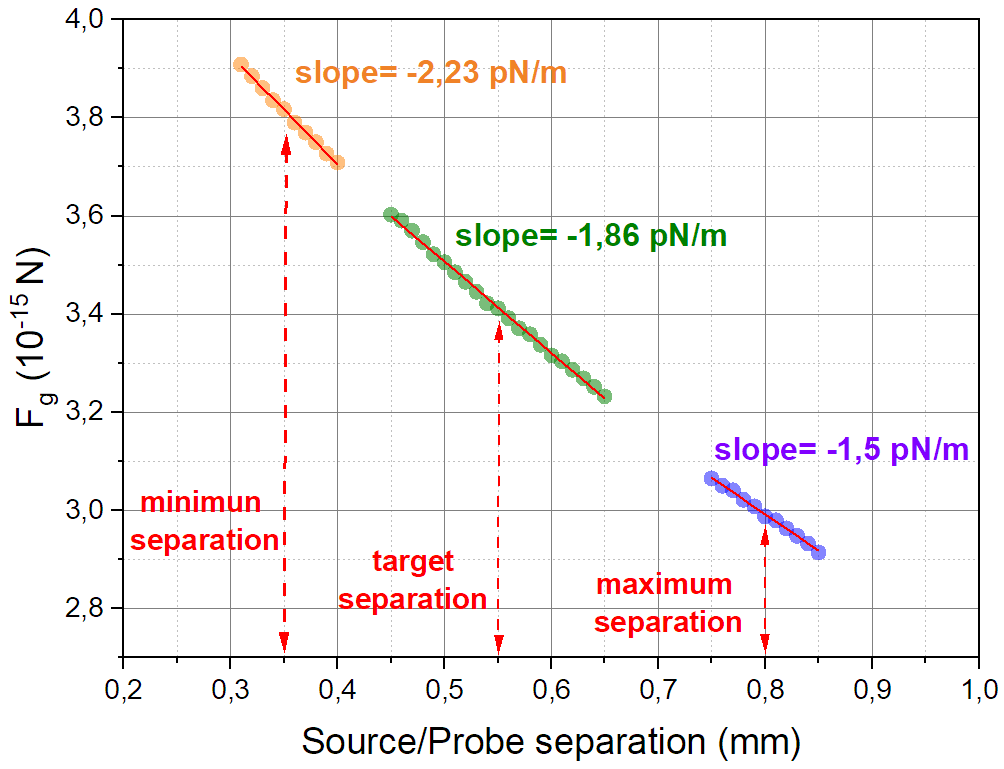}
\caption{Static FEM simulation of the gravitational force as a function of the separation between the Source and Probe masses. The simulation considers the target separation of $550\, \mu$m, together with a conservative and a demanding configuration at $800\, \mu$m and $350\, \mu$m, respectively. Around the nominal separation, $F_{\mathrm{g}}$ is approximately linear, with a slope of $-1.86\times 10^{-12}\,\mathrm{N/m}$; this slope varies by approximately $\pm 20 \%$ over the considered range.\label{fig9}}
\end{figure}

As introduced in Section \ref{sec: Experimental concept}, the experiment, in all its stages, is based on resonantly driving the Source resonator and detecting the resulting displacement of the Probe resonator induced by their mutual gravitational interaction. In this framework, the motion of the Source generates a time-dependent gravitational force expressed in the form: $F(t)=A \,\textrm{Sin}(\omega_{0}t)$, with the amplitude $A$  expressed as $\frac{\partial F_{\mathrm{g}}}{\partial x} \Delta x$, where  $F_{\mathrm{g}}$ is the gravitational force exerted by the Source oscillator on the Probe oscillator, $x$ denotes the coordinate that describes the motion of the normal mode of the Source, and $\Delta x$ is its oscillation  amplitude. To obtain an accurate estimate of the gravitational signal, it is therefore necessary to evaluate $F_{\mathrm{g}}(x)$. In a rigorous treatment, this requires integrating the gravitational interaction over the mass distributions of both resonators, weighted by the modal shapes, thereby capturing the effective coupling associated with the selected driven mode. Owing to the nontrivial geometry of the system, an analytical evaluation of the gravitational interaction and its effect on the Probe is not straightforward. However, a reliable approximation can be obtained by observing that, as shown in Fig. \ref{fig_oscillatori}(c), the fundamental mechanical mode of the oscillator is predominantly characterized by the rigid axial motion of the central cylindrical section.
Therefore, a realistic estimate can be obtained through a static FEM simulation around the distance at which the Source and Probe are fixed. In particular, we employed a finite-element electrostatic simulation, taking advantage of the functional analogy between electrostatic and gravitational forces. The simulation was carried out by assigning a unit charge to the volumes occupied by the Source and the Probe, and evaluating the corresponding Coulomb force $F_{\mathrm{C}}$. Then, the force $F_{\mathrm{g}}$ was obtained by scaling the result as follows: $F_{\mathrm{g}}=F_{\mathrm{C}} \frac{G}{k_{0}}M^2$, where $G$ and $k_{0}$ are respectively the gravitational constant and the Coulomb constant, and $M$ is the mass of the Source and Probe. The result is reported in Fig. \ref{fig9} as a function of the separation between the main masses of the Source and the Probe. The target separation between the Source and Probe is set to $550 \,\mu$m. However, in this analysis, we also consider two additional scenarios: a more conservative configuration with a separation of $800\,\mu $m, and a more demanding one with a separation of $350\,\mu$m, corresponding to the minimum achievable distance given the constraints of the experimental setup. As expected, for small variations around the nominal separation, $F_{\mathrm{g}}$ exhibits an approximately linear dependence, with a slope of $-1.86\times 10^{-12}\,\mathrm{N/m}$ at the target distance and a variation of about $\pm 20 \%$ when the separation is varied across the two extreme values shown in the figure.

We consider driving the Source oscillator with an amplitude corresponding to a displacement of its main mass in the range $(1 \text{-} 5)\, \mu$m. The lower bound is consistent with values already demonstrated in silicon oscillators of comparable mass and frequency \cite{Bawaj2015}, while the upper bound corresponds to a maximum stress of approximately 
10 MPa, providing a safety factor of 5 relative to the fracture limit. Under these conditions, the gravitational force acting on the Probe, located at a nominal distance of $550\, \mu$m from the Source, has an amplitude in the range of 
$(1.86 \,\text{--}\, 9.3)\times10^{-18}\,$N. This has to be compared with the thermal force noise amplitude spectral density, as evaluated from equation \eqref{thermal_force_PSD}, of $4.5 \times 10^{-16}\, \mathrm{N/\sqrt{Hz}}$. The Probe's displacement amplitude at resonance, derived from the transfer function in Fig.~\ref{fig_TF}, falls in the range $(8\text{--}40)\times 10^{-17}\,\mathrm{m}$.

\subsection{Signal detection}
\label{Signal_detection}
The gravitational signal must be compared to a force noise background dominated by thermal fluctuations. Additional contributions to the force noise arise from measurement back-action and possible acoustic disturbances, however we neglect them in the present preliminary estimate. Both signal and noise force are transduced into displacement via the effective susceptibility of the Probe oscillator, which may deviate from its intrinsic form due to cavity optomechanical effects (optical spring and optical damping), as well as possible optomechanical feedback.
We emphasize that optomechanical effects, although modifying the effective susceptibility and inducing cooling of the oscillator, do not alter either the Signal-to-Noise Ratio (SNR) or the useful measurement bandwidth. The optical spring effect in the bad-cavity regime (cavity linewidth exceeding the mechanical resonance frequency) may nonetheless be exploited to slightly tune the resonance frequency.

We assess the measurement sensitivity over the effective bandwidth, focusing in particular on operation at the Probe resonance under phase-sensitive detection. The noise variance can be expressed as $\sigma^2 = S_{xx}/(2\pi \tau)$, where $\tau$ denotes the integration time. The signal, on the other hand, has a root-mean-square value equal to $A_{\mathrm{P}}^2/2$, where $A_{\mathrm{P}}$ is the zero-to-peak oscillation amplitude. Imposing a unit SNR yields an integration time
$\tau = S_{xx}/\pi A_{\mathrm{P}}^2$,
which is determined by the attainable amplitude of the gravitational force signal.
Based on the experimental parameters reported in the previous sections, a unit SNR is expected to be achieved with an integration time ranging from a few minutes to approximately 15 hours, depending primarily on the range of oscillation amplitudes considered for the Source mass. This indicates that the experiment can operate under experimentally accessible conditions.

\subsection{Spurious signals}
\label{sec:Spurious}
Suppressing spurious signals that could obscure the gravitational interaction is of primary importance. The described experimental configuration specifically mitigates two main sources of systematic error: electrostatic coupling between the oscillators and direct mechanical cross-talk.

Concerning electrostatic effects, estimates indicate that, for the adopted oscillator design and in the absence of the electrostatic shield, a parasitic signal comparable to the gravitational one could originate from residual charges of the order of $10^3$ elementary charges, localized on the surfaces of both oscillators, a configuration corresponding to a worst-case scenario. It is known \cite{Krick1988} that SiN membranes can embed elementary charges, and preliminary measurements have demonstrated an overall presence of $10^3$--$10^5$ elementary charges on our manufactured membranes. For this reason as well, a key component of the apparatus is the metallic coating applied to the bifunctional element positioned between the oscillators, which acts to shield electrostatic interactions. The residual force due to electrostatic interaction will be characterized with particular attention, both to exclude the effect of possible electrostatic patch potentials \cite{antonucci2012}  and to evaluate the necessity of controlling the membrane charge state via exposure to UV radiation.

Regarding direct mechanical coupling, we observe that  the target oscillation amplitude of the Source is $1\,\mu$m  while the oscillation amplitude of the Probe induced by the variable gravitational field is approximately $10^{-16}\,$m. Therefore, the ratio between the Source oscillation amplitude and the Probe response expected from gravitational interaction is approximately ten orders of magnitude under resonant excitation. FEM simulations show that the isolation structures integrated within the oscillators provide amplitude attenuation at the level of eight orders of magnitude. A further suppression of about three orders of magnitude is achieved through the two-stage isolation implemented in the Invar support structure housing the oscillators (Fig. \ref{fig_cavita}). We observe that systems with these levels of overall attenuation, although certainly challenging, have already been implemented in experiments for gravitational wave detection \cite{Bignotto2005}.

\section{Conclusions}

We have presented the design and sensitivity analysis of an optomechanical experiment aimed at detecting the gravitational interaction between two microfabricated milligram-scale oscillators operating at ultra-cryogenic temperature. The proposed apparatus combines several ingredients that have not previously been realized within a single platform: milligram test masses, mechanical resonance frequencies in the tens-of-kilohertz range, ultra-high mechanical quality factors, cavity-enhanced displacement readout, and compatibility with operation close to the quantum regime.

Finite-element modelling and noise analysis indicate that a periodically modulated gravitational force generated by a resonantly driven Source oscillator can be detected by a Probe oscillator with integration times ranging from minutes to hours, depending on the achievable drive amplitude and oscillator separation. The predicted displacement signals, of the order of $10^{-16}\,\mathrm{m}$, are within the reach of state-of-the-art cryogenic optomechanical readout systems.

Beyond the direct observation of gravitational coupling, the experiment provides a realistic pathway toward operation in a regime where the Probe motion is dominated by quantum fluctuations. In this configuration, the gravitational signal remains generated by the coherent motion of a classical Source oscillator, but emerges from a background approaching the quantum limit of the Probe. This constitutes an important intermediate step between purely classical measurements of gravity and future experiments involving gravitational interactions between fully quantum mechanical systems.

More generally, the platform developed here establishes a versatile testbed for investigating weak-force sensing, macroscopic quantum optomechanics, and gravity at short distances. Future upgrades, including ground-state cooling of both oscillators and enhanced control of systematic effects, could enable the exploration of genuinely quantum aspects of the gravitational interaction and contribute to experimental efforts addressing the quantum nature of gravity \cite{Bose2025,Marletto2025,Li_2025,Mari2026}.

\section{Acknowledgments}

This work is supported by INFN (GRAFIQO project).
\newpage

\bibliographystyle{nicebib}
\bibliography{database}

\end{document}